\documentclass[10pt,a4paper,twoside]{article}
\usepackage{epsfig}
\usepackage{baltlat6}
\usepackage{array}
\usepackage{here}
\begin{document}
\ \
\vspace{0.5mm}
\setcounter{page}{495}
\vspace{8mm}

\titlehead{Baltic Astronomy, vol.\,20, 495--502, 2011}

\titleb{STARK BROADENING AND WHITE DWARFS}

\begin{authorl}
\authorb{Milan S. Dimitrijevi\'c}{1,2},
\authorb{Andjelka Kova\v cevi\'c}{3},
\authorb{Zoran Simi\'c}{1} \\ and
\authorb{Sylvie Sahal-Br\'echot}{2}
\end{authorl}

\begin{addressl}
\addressb{1}{Astronomical Observatory, Volgina 7, 11060
Belgrade 38, Serbia;\\
mdimitrijevic@aob.bg.ac.rs, zsimic@aob.bg.ac.rs}
\addressb{2}{Observatoire de Paris, LERMA, 5 Place Jules Janssen, 92190 Meudon, France; sylvie.sahal-brechot@obspm.fr}
\addressb{3}{Department of Astronomy, Faculty of Mathematics, Studentski Trg 15, 11000  Belgrade, Serbia; andjelka@matf.bg.ac.rs}
\end{addressl}

\submitb{Received: 2011 August 8; accepted: 2011 August 15}

\begin{summary}
White dwarf and pre-white dwarf atmospheres are one of the best examples
for the application of Stark broadening research results in astrophysics, due to
plasma conditions very favorable for this line broadening mechanism. For example
in hot hydrogen-deficient (pre-) white dwarf stars $T_{eff}$ = 75 000 K - 180 000 K
and log g = 5.5-8 [cgs]. Even for much cooler DA and DB white dwarfs with typical
effective temperatures of 10 000 K - 20 000 K, Stark broadening is usually the dominant
broadening mechanism.

In this review, Stark broadening in white dwarf spectra is considered and the attention is drawn to the
STARK-B database (http://stark-b.obspm.fr/), containing Stark broadening parameters needed for white dwarf
spectra analysis and synthesis, as well as to the new
search facilities which will provide the collective effort to develop Virtual Atomic and Molecular
Data Center (VAMDC - http://vamdc.org/).

\end{summary}

\begin{keywords}
stars: white dwarfs -- Stark broadening, line profiles -- databases
\end{keywords}

\resthead {Stark broadening and white dwarfs}
{M. S. Dimitrijevi\'c, A. Kova\v cevi\'c, Z. Simi\'c, S. Sahal-Br\'echot}

\sectionb{1}{INTRODUCTION}

Stellar spectroscopy is a powerful tool for investigation of stellar plasma since
by analyzing  stellar spectral lines, we can determine for example
the temperature in particular atmospheric layers, the chemical composition
of stellar plasma, surface gravity, spectral type and
effective temperature. For such purposes, in a number of cases, the influence of
collisions with charged particles on emitting/absorbing atoms and ions is important in astrophysical plasmas,
which results in the broadening of spectral lines, so called Stark broadening.

Plasma conditions in
astrophysical plasmas are exceptionally various, so that Stark
broadening is of interest in plasmas from such extreme
conditions like in the interstellar molecular clouds, where typical electron temperatures are around
30 K or smaller, and typical electron densities 2-15 cm$^{ - 3}$ (see e.g. Dimitrijevi\'c, 2010), up to
electron densities of $Ne=10^{22}-10^{24}$ cm$^{-3}$ and temperatures of $T=10^{6}-10^7$ K.
Particularly favorable conditions for Stark broadening are white dwarf- and  pre white
dwarf - atmospheres, where this broadening mechanism is usually dominant in comparison with a concurrent one - thermal
Doppler broadening.

In this work, we will consider Stark broadening in the impact approximation in white dwarf spectra and the corresponding results obtained by members of the
Group for Astrophysical Spectroscopy on Belgrade Astronomical Observatory, and their partners from France, Tunisia, Russia and Canada. Also, the attention will be drawn to the
STARK-B database (http://stark-b.obspm.fr/), containing Stark broadening parameters needed for white dwarf
spectra analysis and synthesis, as well as to the new
search facilities which will provide the collective effort to develop Virtual Atomic and Molecular
Data Center (VAMDC - http://vamdc.org/ - Dubernet et al. 2010, Rixon et al. 2011).

\sectionb{2}{STELLAR PLASMA RESEARCH AND STARK BROADENING}

As an example of the application of Stark broadening for astrophysical plasma research, we will draw attention that line profiles
enter the modeling of stellar atmospheric layers when we
determine  quantities such as the absorption coefficient and the optical depth $\tau_{\nu}$.
Let we take the direction of gravity as z-direction, dealing with a
stellar atmosphere. If the atmosphere is in macroscopic mechanical
equilibrium and with $\rho$ is denoted gas
density, the optical depth
is

$$\tau_{\nu} = \int_z^\infty \kappa_{\nu}\rho dz,\eqno(1)$$

$$\kappa_{\nu} = N(A,i)\phi_{\nu}{{\pi e^2}\over{mc}}f_{ij}, \eqno(2)$$

\noindent where $\kappa_{\nu}$ is the absorption coefficient at a
frequency $\nu$, $N(A,i)$ is the volume density of radiators in the state
$i$, $f_{ij}$ is the absorption oscillator strength, $m$ is the
electron mass and $\phi_{\nu}$ spectral line profile.

Stark broadening may be also important for radiative transfer and opacity calculations, abundances, surface gravity and chemical composition determination,
spectra analysis, interpretation  and synthesis and astrophysical plasma modeling. In the following, we will give several examples of the investigations of Stark broadening in stellar plasma,
performed in the Group for Astrophysical Spectroscopy on Belgrade Astronomical Observatory.

In a number of papers, the influence of Stark broadening on Au II
(Popovi\'c et al. (1999b), Zr II and Zr III  (Popovi\'c et al. 2001a), Nd II (Popovi\'c et al. 2001b), Co III (Tankosi\'c et al. 2003), Ge I (Dimitrijevi\'c et al. 2003a), Si I (Dimitrijevi\'c et al. 2003b), Ga I (Dimitrijevi\'c et al. 2004), Cd I (Simi\'c et al. 2005), Cr II (Dimitrijevi\'c et al. 2007), and Te I (Simi\'c et al. 2009) spectral lines was considered in the spectra of chemically
peculiar A type stars, and in each case
deeper atmospheric layers are found where the
contribution of this broadening mechanism is dominant or could not
be neglected in comparison with Doppler broadening.
In order to provide the corresponding Stark broadening data for rare-earth peak in elemental abundances distribution, Popovi\'c et al. (1999a) considered  the influence of collisions with charged particles on rare earth ion lines in Ap star atmospheres (La II, La III, Eu II and Eu
III) and  found that errors in
equivalent width synthesis and corresponding abundance
determination may be important if it is not taken into account.

\begin{figure*}[t!]
\centerline{\includegraphics[width=8cm]{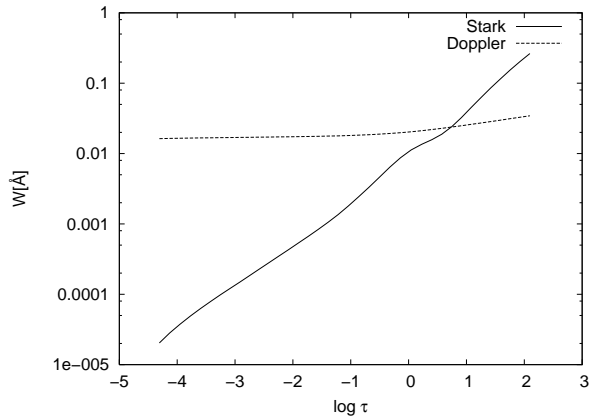}}
\caption{\footnotesize
Thermal Doppler and Stark widths for Te I 6s $^5$S$^o$ - 7p
$^5$P (5125.2 \AA) multiplet as functions of Rosseland optical depth for an A
type star ($T_{eff}$ = 10000 K, log $g$ = 4.5).}
\label{fig1}
\end{figure*}

In order to demonstrate the influence of Stark broadening in atmospheres of hot stars, Fig. 1 shows Stark  widths for Te I 6s $^5$S$^o$ - 7p $^5$P (5125.2 \AA)
multiplet (Simi\'c et al. 2009)  compared with thermal Doppler widths for a model
(T$_{eff}$=10000 K, log g= 4.5) of A type star atmosphere (Kurucz 1979). Namely
thermal Doppler broadening is in stellar atmospheres a not negligible broadening
mechanism and the comparison of Stark and thermal Doppler widths may give an estimate on
the importance of this effect. We note however, that
due to differences in Gauss distribution function for Doppler profile and
Lorentz distribution for Stark, even when Stark width is smaller, in line wings this broadening mechanism may be significant. One can see in Fig. 1,
that in deeper atmospheric layers Stark broadening mechanism becomes first comparable and than dominates in comparison with
thermal Doppler broadening.

\begin{figure*}[t!]
\centerline{\includegraphics[width=8cm]{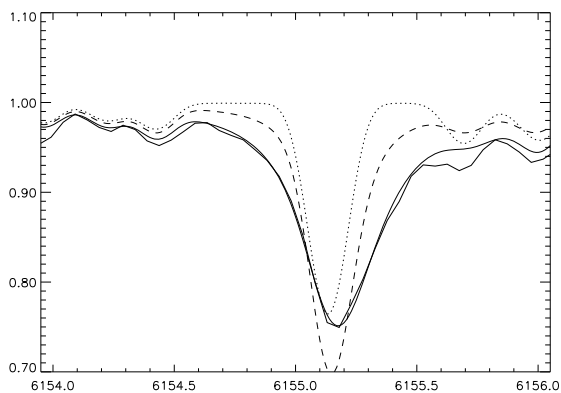}}
\caption{\footnotesize
A comparison between the observed Si I, 6155 \AA\ line profile
in the spectrum of Ap star 10 Aql (thick line) and synthetic spectra
calculated with Stark widths and shifts from Table 1 in Dimitrijevi\'c et al.
(2003b) and Si abundance
stratification (thin line), with the same Stark parameters but for homogeneous
Si distribution (dashed line), and with Stark width calculated by approximation
formulae for the same stratification (dotted line).}
\label{fig2}
\end{figure*}

The influence of Stark broadening and
stratification on neutral silicon lines in spectra of normal late type A star
HD 32115, and Ap stars HD 122970 and 10 Aql was investigated in
Dimitrijevi\'c et al.  (2003b). It was demonstrated that the
synthetic profile of $\lambda $ = 6155.13 {\AA} Si I line fits much better
in the observed one when  Stark broadening is included.

\sectionb{3}{WHITE DWARFS}

From the aspect of Stark broadening in stellar plasmas, white dwarfs are of particular interest. They are at the final
stage of stellar evolution. The first one, 40 Eridani, was discovered by W. Herschel, 31$^{st}$ of March 1783.
The second,  Sirius B, the existence of which was predicted by Bessel in 1844, was discovered by Alvan Graham Clarck in 1862; in 1950 one hundred were known, in 1999 about 2000 and in the Sloan Digital Sky Survey more than 9000.

The traditional division is into hydrogen-rich (DA) and helium-rich (DB), where "D" denotes degenerate objects. Due to gravitational separation in the absence of macroscopic motions  like stellar wind or convection, the most abundant element dominates by several orders of magnitude (Koester 2010) with rare exceptions. Namely while the heavier elements diffuse downward, the lightest ones rest floating on the top of the atmosphere. The origin of H-deficient stars is the late helium shell flash in which white dwarf or post Asymptotic Giant Branch - AGB star reignites helium shell burning, eliminating the rest of hydrogen in this violent event. Post AGB stars are also of great interest from the point of view of need and applicability of Stark broadening data. Namely, AGB are stars which terminated  hydrogen and helium but not carbon burning. They form a sequence of bright red giants - AGB, and they are more luminous than Red Giant Branch stars, which have electron-degenerate helium cores. We note that AGB is often divided in AGB stars with carbon-oxygen cores and Super AGB - SAGB stars with heavier cores.

White dwarfs of DA and DB type have effective temperatures between around 10
000 K and 40 000 K so that Stark broadening is of interest for their spectra
investigation and plasma research, analysis and modeling. When the effective temperature of a cooling white dwarf becomes so low that neither helium nor hydrogen lines are present and only continuum appears in the spectrum, the white dwarf become of DC type. If in the spectra of hydrogen-rich or helium-rich atmospheres appear and lines of various metals (not counting carbon lines which will be discussed later), white dwarfs are of DZ, DAZ or DBZ type. Such metals should be accreted from outside, from interstellar matter or from some tidally disrupted asteroid (Koester 2010).

DB White dwarfs, the most interesting from the point of view of Stark broadening of non-hydrogenic lines,  are divided in the following types: DO, with   40 000 K $< T_{eff} <$ 100000 K (120 000 K according to Dreizler and Werner (1996) and Stark broadening is very important for the investigation of their spectra, Hamdi et al. 2008), with He II  lines in the spectrum, DB, with 12 000 K $< T_{eff} <$ 40000 K, and with  He I lines, and DQ, with 4 000 K $< T_{eff} <$ 12000 K, with C lines and C$_2$ Swan band in the spectrum. It is assumed that carbon in the DQ white dwarfs is extracted from deeper layers by the growing convective zone (Koester 2010).

McGraw et al. (1979) discovered in the 1979, the first object of a new class of hot hydrogen deficient degenerate stars PG1159-035.
In the atmospheres of PG1159 stars, a mixture of helium, carbon and oxygen is present. They most likely experienced a very late thermal pulse which  eliminated the rest of hydrogen, mixed helium with carbon, oxygen and other elements from the envelope and returned them back from white dwarf phase to the post AGB phase.
PG1195 objects are very interesting for applications of results of Stark broadening investigations, with
effective temperatures ranging from $T_{eff}$ = 100 000 K to 140 000 K, where
of course Stark broadening is very important (Werner et al. 1991). They have
high surface gravity (log $g $= 7), and their photospheres are dominated by
helium and carbon with a significant amount of oxygen (C/He = 0.5 and
O/He = 0.13) (Werner et al. 1991). Their spectra, strongly influenced by Stark
broadening, are dominated by He II, C IV, O VI and N V lines.

Nugent et al. (1983) discovered in 1983 H 1504+65, a PG 1195 faint blue star which is not only hydrogen-, but also helium-deficient, with  atmosphere of carbon and oxygen in equal amount.
This is the hottest known star with $T_{eff}$ of 170000 K $\pm $ 20000 K, or "bare core of the former AGB star" according to Werner et Wolff (1999).

The need for the large number of reliable Stark broadening data for a number of trace elements, was highly stimulated  by
FUSE – Far Ultraviolet Spectroscopic Explorer satellite, which provided astronomers with a great number of high resolution spectra of hot evolved stars within the wavelength range 907-1187 \AA. Fontaine et al. (2008) note that FUSE range includes "high density of transitions associated with numerous ionization levels of several elements such as:
C, N, O, Si, S, P, Cl, Ne, Ar, V, Mn, Cr, Fe, Co, Ni, Ge, As, Se, Zr, Te, I and Pb among others."
For analysis and synthesis of spectra and modeling of atmospheres of hot white dwarfs, PG 1159 stars, hot B subdwarfs, post AGB (Asymptotic Giant Branch) objects such as CSPNs (Central Stars of Planetary Nebulae), observed by FUSE, accurate Strak broadening data for great number of spectral lines of various atoms and their ions are needed. However, as it was pointed out in Rauch et al. (2007), line broadening data fore many species and their ions are missing in the literature. Moreover, some existing data are provided within insufficient temperature and density ranges, and extrapolation to the plasma conditions in line forming regions introduces additional errors.

A new type of hot DQ white dwarfs with $T_{eff}$ 18000 - 24000 K, and carbon atmospheres, has recently been discovered by Dufour et al. (2007, 2008), enabling them to propose an evolutionary sequence of white dwarfs from H 1504+65 and PG 1195 objects to hot DQ, DQ and DB white dwarfs. In the newly discovered class of these hot DQ white dwarfs, hydrogen and helium are absent in the atmosphere and dominate lines of C II, while O II lines are also present (Dufour 2011). For the investigation and modeling of these stars, and in order to understand their origin and evolution, the accurate
determination of surface gravity is essential. However, the interpretation and analysis of these
surprising newly discovered objects was humped by the lack of the corresponding Stark broadening data, which additionally illustrates their crucial need for white dwarf atmosphere research and modeling.

Members of Group for Astrophysical Spectroscopy  considered the influence of Stark broadening for DA, DB and DO white dwarf
atmospheres (Popovi\'c et al. 1999b,Tankosi\'c et al. 2003, Milovanovi\'c et al. 2004, Simi\'c et al. 2006, Hamdi et al. 2008) and it was found that for such stars
Stark broadening is dominant in practically all relevant atmospheric layers.

As an example, Simi\'c et al. (2006) considered the influence of Stark broadening on Cu III, Zn III and Se III
spectral lines in DB white dwarf atmospheres  for 4s $^2$F - 4p $^2$G$^o$
($\lambda$=1774.4 \AA), 4s $^3$D - 4p $^3$P$^o$ ($\lambda$=1667.9 \AA)
and 4p5s $^3$P$^o$ - 5p $^3$D ($\lambda$=3815.5 \AA)
by using the corresponding model  with
$T_{eff}$ = 15000 K and $\log g$ = 7 (Wickramasinghe 1972). For this analysis,
optical depth points at the standard wavelength $\lambda_{s}$=5150
\AA ($\tau_{5150}$) are used as in Wickramasinghe (1972). One can see in Fig. 3 that the
thermal Doppler broadening has a much less importance in comparison
with the Stark broadening mechanism. If one compares Fig. 1 and Fig. 3, one can see that in comparison with A type stars, the importance of Stark broadening  is much greater for DB white dwarf atmospheres. This is the consequence of larger electron densities due to  larger
$\log g$ and  $T_{eff}$, so that electron-impacts producing Stark broadening are more effective.

\begin{figure*}[t!]
\centerline{\includegraphics[width=8cm]{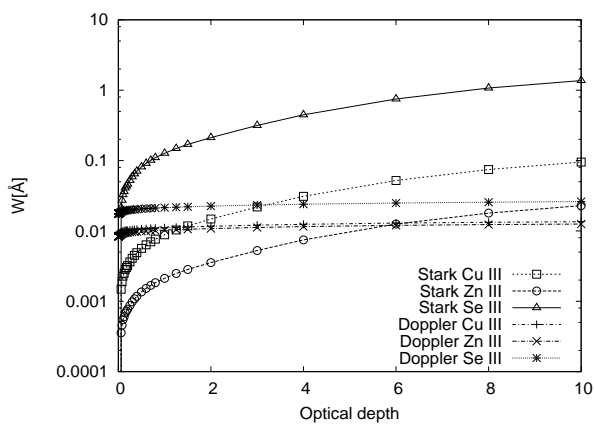}}
\caption{\footnotesize
Thermal Doppler and Stark widths for
Cu III 4s $^2$F - 4p $^2$G$^o$ ($\lambda$=1774.4 \AA), Zn III
4s $^3$D - 4p $^3$P$^o$ ($\lambda$=1667.9 \AA) and Se
III 4p5s $^3$P$^o$ - 5p $^3$D ($\lambda$=3815.5 \AA)
spectral lines for a DB white dwarf atmosphere
model with $T_{eff}$  = 15,000 K and $log$ g = 7, as a function of
optical depth ${\tau}_{5150}$.}
\label{fig3}
\end{figure*}

\begin{figure*}[]
\centerline{\includegraphics[width=8cm]{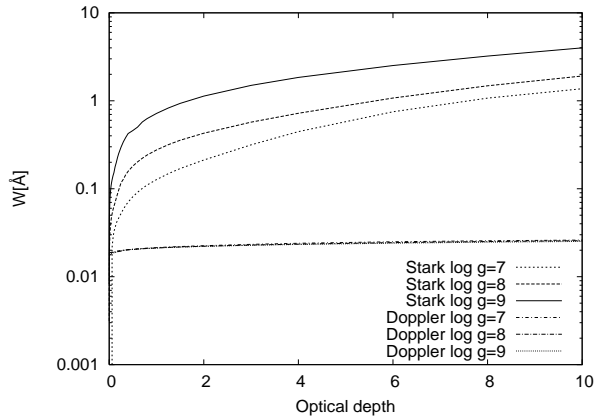}}
\caption{ Thermal Doppler and Stark widths for Se III spectral line
5s $^3$P$^o$ - 5p $^3$D ($\lambda$=3815.5 \AA) for a DB white dwarf
atmosphere model with $T_{eff}$  = 15000 K and 7 $\leq$ $log$ g
$\leq$ 9, as a function of optical depth ${\tau}_{5150}$.}
\end{figure*}

For example is shown in Fig. 4 that the Stark width of the
Se III 3815.5 \AA\ line is larger than the Doppler one by up to
two orders of magnitude within the range of optical
depths considered.

Hamdi et al. (2008) considered the broadening on Si VI
lines in DO white dwarf spectra for 50000 K $\leq$ $T_{eff}$ $\leq$
100000 K and for 6 $\leq$ log $g$ $\leq$ 9 atmosphere models. They found that the
influence of Stark broadening  increases with log $g$ and is dominant in broad
regions of the considered DO white dwarf atmospheres.

Currently, within a wide international collaboration (Canada, France, Tunisia, Serbia), the work on the Stark broadening of carbon and oxygen lines in hot DQ white dwarf spectra (Dufour et al. 2011) is in progress.

Reliable Stark broadening data for white dwarf research may be found in STARK-B database (http://stark-b.obspm.fr), dedicated for modeling of stellar atmospheres, stellar spectra analysis and synthesis,
as well as for laboratory plasma research, inertial fusion plasma, laser development and investigations of plasmas in various technologies.
This database enters also in FP7 project Virtual Atomic and Molecular Data Centre - VAMDC (P.I. Marie Lise Dubernet), an European FP7 project with aims: (i) to build a secure, flexible and interoperable e-science environment based interface to the existing Atomic and Molecular  databases; (ii) to coordinate groups involved in the generation, evaluation, and use of atomic and molecular data, and (iii) to  provide a forum for training of potential users (Dubernet et al. 2010, Rixon et al. 2011)  .

In future, a mirror site will be a part of SerVO - Serbian Virtual Observatory (http://www.servo.aob.rs/$\sim$darko - P.I. Darko Jevremovi\'c), which main goals are to publish in VO compatible format data obtained by Serbian astronomers, in particular digitized photographic plates from the Belgrade Astronomical Observatory archive, in order to make them accessible to scientific community, as well as to provide astronomers in Serbia with VO tools for their research (Jevremovi\'c et al. 2009)

\thanks{ This work has been supported by VAMDC,  funded under the "Combination of Collaborative Projects and Coordination and  Support Actions" Funding Scheme of The Seventh Framework Program. Call topic: INFRA-2008-1.2.2 Scientific Data Infrastructure. Grant Agreement number: \break
239108.  The authors are also  grateful for the support  provided by Ministry Education and Science of Republic of Serbia through project  176002 "Influence of collisional processes on astrophysical plasma spectra", and III44002 "Astroinformatics: Application of IT in astronomy and related disciplines".}

\References

\refb Dimitrijevi\'c M. S. 2010, Mem.S.A.It., 75, 282

\refb Dimitrijevi\'c M. S., Da\v ci\' c M., Cvetkovi\' c Z. Simi\' c Z., 2004, A\&A 425, 1147

\refb Dimitrijevi\'c M. S., Jovanovi\' c P. Simi\' c Z., 2003a, A\&A 410, 735

\refb Dimitrijevi\'c M. S., Ryabchikova T., Popovi\'c L. \v C., Shylyak D. and Tsymbal V., 2003b, A\&A 404, 1099

\refb Dimitrijevi\'c M.S., Ryabchikova T., Simi\' c Z., Popovi\'c L. \v C. and Da\v ci\' c M., 2007, A\&A 469, 681

\refb
Dreizler S. and Werner K., 1996, A\&A 314, 217

\refb Dubernet M. L., Boudon V., Culhane J. L. et al., 2010, JQSRT  111, 2151

\refb Dufour P., Ben Nessib N., Sahal-Br\'echot S., Dimitrijevi\'c M. S., 2011, Baltic Astron. this issue

\refb Dufour P., Fontaine G., Liebert J., Schmidt G. D., Behara N., 2008, ApJ 683, 978

\refb  Dufour P., Liebert J., Fontaine G., Behara N.,  2007, Nature Letters, 450, 522

\refb Fontaine M., Chayer P., Oliveira C. M. et al., 2008, ApJ 678, 394

\refb Hamdi R., Ben Nessib N., Milovanovi\' c N., Popovi\'c  L. \v C ., Dimitrijevi\'c M. S. and Sahal-Br\'echot S., 2008, MNRAS 387, 871

\refb
Jevremovi\'c D., Dimitrijevi\'c M. S., Popovi\'c L. \v{C}. et al.,  2009, New Astron. Rev. 53, 222

\refb Koester D., 2010, Mem. S. A. It. 921

\refb
Kurucz R. L., 1979, ApJS 40, 1

\refb McGraw J. T., Starrfield S., Liebert J., Green R. F., 1979, Proc. IAU Coll. 53: White dwarfs and variable degenerate stars, H. M. van Horn, V. Weidemann eds., Univ. of Rochester, 377

\refb Milovanovi\' c N., Dimitrijevi\'c M. S., Popovi\'c  L. \v C., Simi\' c Z., 2004, A\&A 417, 375

\refb Nugent J. J., Jensen K. A., Nousek J. A. et al. 1983, ApJS, 51, 1

\refb  Popovi\'c L.\v C., Dimitrijevi\'c M. S. and Ryabchikova T., 1999a, A\&A 350, 719

\refb  Popovi\'c L.\v C., Dimitrijevi\'c M. S. and Tankosi\' c D., 1999b, A\&AS 139, 617

\refb  Popovi\'c  L.\v C ., Milovanovi\' c N. and Dimitrijevi\'c M. S., 2001a, A\&A 365, 656

\refb  Popovi\'c L.\v C ., Simi\' c S., Milovanovi\' c N., Dimitrijevi\'c M.S., 2001b, ApJS 135, 109

\refb Rauch T., Ziegler M., Werner K., Kruk J. W., Oliveira C. M., Vande Putte D., Mignanin R. P., Kerber F., 2007, A\&A 470, 317

\refb  Rixon G., Dubernet M. L., Piskunov N. et al., 2011, AIP Conference Proceedings 1344, 107

\refb  Simi{\' c} Z., Dimitrijevi\'c M. S. and Kova\v cevi\'c A., 2009, New Astron. Rew. 53, 246

\refb  Simi{\' c} Z., Dimitrijevi\'c M. S., Milovanovi\' c N. and Sahal-Br\'{e}chot, S., 2005, A\&A 441, 391

\refb Simi{\' c} Z., Dimitrijevi\'c M.S., Popovi\'c L.\v C., Da\v ci\' c M., 2006, New Astron. 12, 187

\refb  Tankosi{\' c} D., Popovi\'c L. \v C. and Dimitrijevi\'c M. S., 2003, A\&A 399, 795

\refb
Werner K. and Heber U. and Hunger R., 1991, A\&A 244, 437

\refb
Werner K. and Wolff B., 1999, A\&A 347, L9

\refb
Wickramasinghe, D.T. 1972, Mem. R. Astron. Soc., 76, 129

\end{document}